\documentclass[referee,pdflatex,sn-nature]{sn-jnl}% referee option is meant for double line spacing

\usepackage{graphicx}%
\usepackage{multirow}%
\usepackage{amsmath,amssymb,amsfonts}%
\usepackage{amsthm}%
\usepackage{mathrsfs}%
\usepackage[title]{appendix}%
\usepackage{xcolor}%
\usepackage{textcomp}%
\usepackage{manyfoot}%
\usepackage{booktabs}%
\usepackage{algorithm}%
\usepackage{algorithmicx}%
\usepackage{algpseudocode}%
\usepackage{listings}%
\theoremstyle{thmstyleone}%
\theoremstyle{thmstyletwo}%

\theoremstyle{thmstylethree}%

\newcommand{\hc}[2]{C\textsubscript{#1}H\textsubscript{#2}}

\begin{document}

\title[FPGA-Native Semi-Empirical Electronic Structure Theory]{A Hardware-Native Realisation of Semi-Empirical Electronic Structure Theory on Field-Programmable Gate Arrays}


\author[1]{\fnm{Xincheng} \sur{Miao}}\email{xincheng.miao@uni-wuerzburg.de}

\author*[1]{\fnm{Roland} \sur{Mitri\'{c}}}\email{roland.mitric@uni-wuerzburg.de}

\affil*[1]{%
  \orgdiv{Institut für Physikalische und Theoretische Chemie}, %
  \orgname{Julius-Maximilians-Universit\"at W\"urzburg}, %
  \orgaddress{\street{Emil-Fischer-Straße 42}, \city{W\"urzburg}, %
    \postcode{97074}, %
    %\state{Bavaria}, %
    \country{Germany}%
  }%
}


\abstract{%
    High-throughput quantum-chemical calculations underpin modern
    molecular modelling, materials discovery, and machine-learning workflows, 
    yet even semi-empirical methods become restrictive when many molecules 
    must be evaluated. 
    Here we report the first hardware-native realisation of semi-empirical 
    electronic structure theory on a field-programmable gate array (FPGA),
    implementing as a proof of principle Extended H\"uckel Theory (EHT) 
    and non-self-consistent Density Functional Tight Binding (DFTB0).
    Our design performs Hamiltonian construction and diagonalisation on the 
    FPGA device through a streaming dataflow, enabling deterministic execution without host 
    intervention. On a mid-range Artix-7 FPGA, the DFTB0 Hamiltonian generator delivers 
    a throughput over fourfold higher than that of a contemporary server-class CPU.
    Improvements in eigensolver design, memory capacity, and extensions to 
    nuclear gradients and excited states could further expand capability. 
    Combined with the inherent energy efficiency of FPGA dataflow, 
    this work opens a pathway towards sustainable, hardware-native acceleration 
    of electronic-structure simulation and direct hardware implementations of 
    a broad class of methods.
}




\maketitle

\section{Introduction}\label{sec:introduction}

Quantum electronic structure calculations are central to molecular modelling and materials 
design, since they allow to predict properties from first principles. 
Modern \emph{ab initio} electronic-structure methods offer high accuracy but remain
computationally intensive. This become especially restrictive whenever the same type of
calculations must be repeated for many different molecular structures.
This situation arises routinely in high-throughput materials screening, where large 
libraries or composition/structure spaces are explored to down-select candidates with
targeted properties. Major community efforts (e.g. materials project and
high-throughput workflows) have shown the importance of automated
multi-structure evaluation to accelerate materials
design.~\cite{curtarolo_high-throughput_2013,jain_commentary_2013,luo_highthroughput_2021,luo_highthroughput_2021-1,ren_high-throughput_2022}
Likewise, potential energy surface (PES) exploration for data-driven interatomic models,
as used in modern artificial intelligence based force field development, 
relies on diverse, well curated configurations spanning relevant
regions of chemical space.~\cite{unke_machine_2021,kulichenko_data_2024} 
Similarly, in molecular dynamics, extensive ensembles of geometries are indispensable to
capture kinetics and thermodynamics.~\cite{yang_enhanced_2019,henin_enhanced_2022,mohr_enhanced_2024}
As these workloads scale, the cumulative wall time and energy consumption associated 
with repeated electronic-structure evaluations become increasingly significant.
For \emph{ab initio} methods, much of this cost is attributed to the evaluation and manipulation of 
electron-repulsion integrals, which often dominate memory traffic and arithmetic in many 
computational workflows. 
Numerous algorithmic advances have substantially reduced their cost,
\cite{gill_simple_1994,reine_multielectron_2012,sandberg_algorithm_2012,sun_libcint_2015,zhang_libreta_2018,holzer_improved_2020,laqua_accelerating_2021,helmich-paris_improved_2021,asadchev_3-center_2024,zhang_accelerating_2025}
yet integral handling remains a major performance bottleneck in \emph{ab initio}
calculations. 
Beyond mean-field theory, correlated wave-function approaches introduce very large 
configuration spaces, which further amplify computational effort,
leading ultimately to an exponential scaling in the exact limit.
As a result, despite decades of algorithmic optimisation, the cost of 
electronic-structure evaluation continues to limit the size of systems and the 
number of configurations that can be explored in practice.

To address this bottleneck, semi-empirical electronic structure methods 
offer a pragmatic accuracy–cost compromise 
for such multi-geometry workloads: they retain a quantum Hamiltonian structure while
replacing the most expensive \emph{ab initio} components with parametrised
terms in the Hamiltonian, enabling both routine MD and high-throughput studies 
at a fraction of the cost of \emph{ab initio} methods.~\cite{hoffmann_extended_1963,porezag_construction_1995,elstner_self-consistent-charge_1998,spiegelman_density-functional_2020}
Nevertheless, large numbers of semi-empirical calculations can still demand considerable 
computational resources on conventional CPU-centric workflows. This motivates approaches 
that increase throughput and lower the cost per calculation, through 
algorithmic improvements and/or alternative computing architectures, 
while reducing the energy footprint of large simulation campaigns.
\cite{omar_high-throughput_2021,bosia_ultra-fast_2023}

One established response is to run electronic structure calculations on
graphical processing units (GPUs).~\cite{walker_electronic_2016} 
In such workflows, the compute-intensive parts 
of the calculation are executed on the GPU, while the CPU mainly handles control. 
Examples for \emph{ab initio} electronic structure calculations and dynamics simulations 
include implementations in
TeraChem~\cite{seritan_terachem_2021},
QUICK~\cite{miao_acceleration_2015,manathunga_quantum_2023},
GPU4PySCF~\cite{li_introducing_2025}, and 
VeloxChem~\cite{li_veloxchem_2025} codes. 
Similarly, for semi-empirical methods, there
are GPU-based implementations, including QUELO-G~\cite{gunst_quantum_2024},
PySEQM~\cite{athavale_pyseqm_2025}, as well as GPU-capable tight-binding
frameworks such as DFTB+~\cite{hourahine_dftb_2020} and
xtb~\cite{bannwarth_extended_2021,steinbach_acceleration_2025}. Taken together,
these examples show that moving core electronic-structure kernels to GPUs can
shorten the time for computing a single geometry and increases overall
throughput when many geometries are processed.

Despite the substantial acceleration achieved on GPUs, conventional workflows
still encounter important limitations. Running very many electronic-structure 
calculations leads to noticeable overhead from repeated kernel
launches and synchronisation.~\cite{kurzak_implementation_2016} Control flows
that depend strongly on the type of orbital shells or screening conditions are
difficult to map efficiently to the single-instruction–multiple-thread (SIMT)
models, reducing effective utilisation of the available parallel hardware.~\cite{tsuji_gpu_2025}
In addition, intermediate data are often written to global memory between kernels, 
increasing traffic and latency time. All these factors motivate the exploration of dedicated 
architectures tailored specifically to electronic-structure calculations, which can be realised 
using field-programmable gate arrays (FPGAs).

Studies have shown that FPGAs can accelerate a variety of chemistry workloads. 
Early efforts demonstrated FPGA acceleration of classical molecular dynamics (MD),~\cite{gu_fpga_2008} with custom 
micro-architectures for force evaluation and on-chip data management yielding significant speedups 
over CPUs. A comparative study investigated numerical 
quadrature for two-electron integrals on FPGAs alongside GPUs and other accelerators, 
highlighting distinct implementation strategies.~\cite{gillan_comparing_2012}
More recent developments reported fully integrated FPGA-based MD engines with on-chip particle storage, 
achieving performance competitive with GPU-based AMBER classical force-field simulations.~\cite{yang_fully_2019} 
FPGA acceleration has also been applied to real-time quantum dynamics, demonstrating fast execution 
with markedly lower energy consumption than conventional compute nodes.~\cite{rodriguez-borbon_field_2020}
Further work introduced FPGA kernels for electron-repulsion integrals (ERIs), exceeding 
twenty billion integrals per second and significantly outperforming optimised CPU and GPU
baselines.~\cite{wu_computing_2023,stachura_seri_2024}
Most recently, FPGA clusters for classical MD interconnected \emph{via} a 3D torus topology have been demonstrated, 
reaching microsecond-per-day simulation rates and showing scalability to larger systems through 
the addition of nodes.~\cite{hamm_toward_2025}
Taken together, these efforts demonstrate how reconfigurable hardware with deeply
pipelined dataflow, fine-grained parallelism, and efficient on-chip memory reuse can be 
applied across chemistry tasks from integrals to dynamics, offering clear potential for 
enabling high-throughput electronic-structure and simulation workloads.

While previous studies have demonstrated the use of FPGAs for specific kernels
and chemistry-related tasks, there has not yet been a hardware-native implementation
of a complete electronic-structure method, in which the numerical workload runs
entirely on the FPGA fabric without assistance from external processors. Such
an approach avoids host–device communication overhead, allows successive
functions (for example, Hamiltonian assembly and diagonalisation) to be fused
into a single streaming pipeline, and provides deterministic execution with
predictable latency.

Here, we present the first hardware-native FPGA realisation of a semi-empirical 
electronic-structure method. For demonstration purposes, we implement both
Extended Hückel Theory (EHT) and non-self-consistent density functional
tight binding (DFTB0) methods by performing Hamiltonian construction and diagonalisation
entirely on the FPGA device. 
In addition, we developed a separate streaming implementation for the 
DFTB0 Hamiltonian construction whose performance rivals that of an optimised CPU code, 
illustrating the potential of on-device pipelines for high-throughput 
semi-empirical calculations. 
This opens an avenue for the implementation of various electronic-structure methods 
directly in hardware, suggesting a path toward FPGA-based realisations of a broad range 
of semi-empirical and \emph{ab initio} approaches.

% The remainder of this paper outlines the FPGA design and implementation of the
% EHT and DFTB0 kernels (Section~\ref{sec:implementation}), presents performance
% benchmarks against CPU references (Section~\ref{sec:numerical_experiments}), and
% discusses prospects for high-throughput electronic-structure calculations
% (Section~\ref{sec:conclusions_and_outlook}).

\section{Results}\label{sec:results}

\subsection{Semi-Empirical Electronic Structure Methods}\label{subsec:semi_empirical_methods}

In the electronic structure theory we commonly consider electronic systems described by 
the non-relativistic Hamiltonian in second quantisation
\begin{equation}
  \hat{H} = \sum_{ij} h_{ij} a_i^\dagger a_j 
  + \frac{1}{2} \sum_{ijkl} V_{ijkl} a_i^\dagger a_j^\dagger a_k a_l\,,
\label{eq:second_quantised_hamiltonian}
\end{equation}
where $h_{ij}$ and $V_{ijkl}$ denote the one-electron and two-electron integrals
in a spin-orbital basis, and $a_i^{(\dagger)}$ denotes the fermionic operators.
Within a single-determinant mean-field approximation 
and a fixed atomic-orbital basis, 
the stationary electronic problem reduces to the generalised eigenvalue equation
\begin{equation}
\boldsymbol{H} \boldsymbol{C} = \boldsymbol{S} \boldsymbol{C} \boldsymbol{\epsilon},
\label{eq:generalized_evp}
\end{equation}
with Hamiltonian matrix $\boldsymbol{H}$ and overlap matrix $\boldsymbol{S}$.
The matrix $\boldsymbol{C}$ contains the molecular-orbital coefficients and
$\boldsymbol{\epsilon}$ is the diagonal matrix of orbital energies.
This is a generic problem whose structure is the same across wide range of 
semi-empirical and \emph{ab initio} electronic structure methods.

For example, in Extended H\"{u}ckel Theory (EHT)~\cite{hoffmann_extended_1963}, the effective
one-electron Hamiltonian is constructed directly from atomic valence orbital
energies and overlap integrals. 
In this work, we adopt the Extended Hückel Neglect of Differential Overlap (EHNDO) formulation~\cite{dixon_fast_1994}, 
in which differential overlap contributions are neglected and 
orbital-specific empirical scaling factors are introduced. 
The resulting Hamiltonian matrix elements are given by
\begin{equation}
H_{\mu\nu} = \begin{cases}
\epsilon_{\mu}, & \mu = \nu, \\
k_\mu k_\nu (\epsilon_{\mu}+\epsilon_{\nu}) S_{\mu\nu}, & \mu \neq \nu,
\end{cases}
\label{eq:eht_hamiltonian_general}
\end{equation}
where $\epsilon_{\mu}$ denotes the valence orbital energy of atomic orbital
$\mu$, $S_{\mu\nu}$ is the overlap integral, and $k_\mu$, $k_\nu$ are empirical
scaling factors.

Within the EHNDO formulation, an overlap-neglect approximation is employed 
for the the generalised eigenvalue problem in Eq.~\eqref{eq:generalized_evp},
such that it reduces to the ordinary eigenvalue equation
\begin{equation}
\boldsymbol{H}\boldsymbol{C}=\boldsymbol{C}\boldsymbol{\epsilon},
\label{eq:standard_evp}
\end{equation}
which is solved to obtain the molecular-orbital energies and coefficients.

The same overlap-neglect approximation was applied also to the 
non-self-consistent Density Functional Tight Binding (DFTB0)
method~\cite{porezag_construction_1995}, where the Hamiltonian matrix elements are constructed from 
precomputed two-centre integrals derived from atomic Kohn–Sham density functional theory calculations. 
These integrals are tabulated as functions of the interatomic distance and combined according to 
Slater--Koster rules~\cite{slater_simplified_1954}, 
which express the matrix elements between oriented atomic orbitals in terms of a small set of 
direction-independent integrals. 
For a given pair of orbitals $\mu$ and $\nu$, the Hamiltonian element can be written as
\begin{equation}
  H_{\mu\nu}(\vec{R}) = \sum_{\alpha} f^{\alpha}_{\mu\nu}(\hat{R}) H^{\alpha}(R)\,,
\label{eq:slater_koster}
\end{equation}
where $R=\|\vec{R}\|$ is the interatomic distance, $\hat{R}=\vec{R}/R$ is the 
corresponding direction vector, $H^{\alpha}(R)$ are the tabulated two-centre matrix elements 
(\emph{e.g.} $\sigma$, $\pi$, or $\delta$ contributions), 
and $f^{\alpha}_{\mu\nu}(\hat{R})$ are the geometry-dependent Slater–Koster coefficients.
DFTB0 additionally includes a short-range repulsive potential between atoms, 
accounting for nuclear–nuclear repulsion and approximating contributions from 
neglected electron–electron interactions.

In this work, as a proof-of-principle demonstration, we implement both methods on an 
FPGA device and explore the computational performance of this device.

\subsection{FPGA Design and Computational Workflow}

\begin{figure}[h]
    \centering
    \hspace*{-2.8cm}
    \includegraphics[scale=1.0]{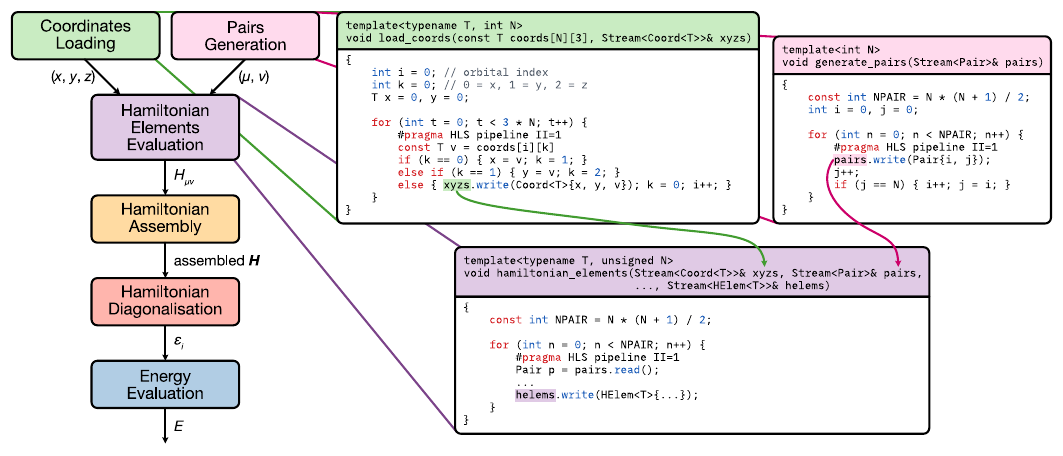}
    \caption{
        Schematic dataflow graph of the tight-binding HLS workflow with kernel-level detail.
        The electronic-structure calculation is mapped to a streaming task graph 
        composed of coordinate loading, pair generation, Hamiltonian-element evaluation, 
        matrix assembly, diagonalisation, and energy evaluation. Independent HLS kernels 
        communicate \emph{via} buffered streams that carry coordinates,
        orbital-index pairs, and Hamiltonian elements. 
        Representative HLS/C++ code excerpts illustrate the corresponding stream interfaces 
        and pipelined loop structure.
        The assembled Hamiltonian enters the diagonalisation stage once the matrix is fully assembled.
    }
    \label{fig:tb_dataflow}
\end{figure}

Our implementation is carried out using the Vitis High-Level Synthesis (HLS) workflow, 
which converts algorithms described in the C/C++ programming language into hardware implementations,
and is deployed on a Xilinx Artix-7 FPGA platform (see Sec.~\ref{subsubsec:devices_tools_clocking}).
At a high level, the tight-binding workflow is realised as a streaming task graph 
composed of coordinate loading, pair generation, Hamiltonian-element evaluation, 
matrix assembly, and diagonalisation (Fig.~\ref{fig:tb_dataflow}). 
These stages execute as independent HLS kernels connected by streaming interfaces, 
allowing data to propagate through the pipeline as soon as it is produced. 
Orbital pairs are therefore processed in a fine-grained, element-wise fashion, 
with downstream stages operating on previously emitted pairs while upstream stages 
advance to subsequent ones. This organisation leads to substantial temporal overlap 
across the workflow and enables continuous utilisation of the computational pipeline.
For illustration, we show HLS/C++ code excerpts in Fig.~\ref{fig:tb_dataflow} that 
highlight the stream interfaces and the pipelined loop structure of the kernels.
The complete source code can be found in the supplementary information.

To support this streaming execution model, nested loops over orbital indices are eliminated 
by generating orbital pairs explicitly in a dedicated pair-generation stage. 
This transforms the original double loops into a flat stream of index pairs that is consumed uniformly
by all downstream kernels, simplifying loop structure and enabling efficient pipelining. 
As a result, all loops processing orbital pairs can be scheduled with an initiation interval of one, 
allowing one Hamiltonian element to be produced per cycle once the pipeline is filled. 
After an initial warm-up period, the inter-geometry spacing is determined by the slowest stage 
in the task graph rather than by the sum of all stage runtimes. 
Efficient use of hardware resources is further supported by employing minimal-width, 
arbitrary-precision data types for indices and addresses.
Further details on loop structuring, pipelining, and use of arbitrary-precision data types 
are provided in Sec.~\ref{subsubsec:general_design_principles}.

The same streaming workflow is used for both Extended H\"{u}ckel Theory (EHT) and 
non-self-consistent DFTB (DFTB0); method-specific differences arise only in the evaluation of 
individual Hamiltonian elements. In EHT, these elements are computed directly from overlap integrals 
and tabulated orbital parameters, whereas in DFTB0 they are obtained from pre-tabulated two-centre integrals 
combined according to Slater--Koster rules. The corresponding hardware kernels differ in their 
internal arithmetic and memory access patterns, but interface identically with the surrounding streaming pipeline. 
Detailed descriptions of the EHT and DFTB0 kernels are given in 
Secs.~\ref{subsubsec:method_specific_eht} and \ref{subsubsec:method_specific_dftb0}.

The repulsive pair potential in DFTB0 is evaluated by a dedicated kernel that operates independently 
of the Hamiltonian evaluation and diagonalisation stages. This contribution is therefore fully overlapped 
with the main workflow, and its computational cost is hidden by the longer-running stages of the pipeline. 
As a result, the repulsive term introduces negligible overhead in the overall runtime. 
Implementation details are provided in Sec.~\ref{subsubsec:method_specific_dftb0}.

\begin{figure}[h]
    \centering
    \hspace*{-2.6cm}
    \includegraphics[scale=1.0]{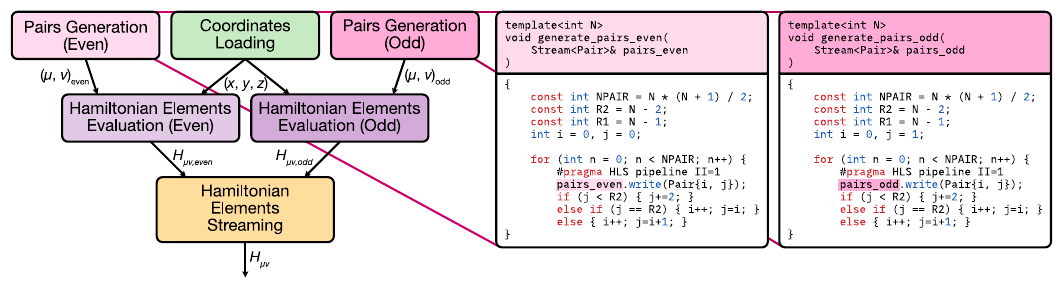}
    \caption{
        Schematic dataflow graph of the stand-alone Hamiltonian-generation HLS workflow with kernel-level detail.
        Pair generation and Hamiltonian-element evaluation are duplicated for even- and odd-indexed orbital pairs to sustain peak throughput. 
        Coordinates are broadcast to both branches, while each branch produces 
        a stream of elements which are subsequently merged by a streaming stage 
        into a single output stream.
        Representative HLS/C++ code excerpts show the corresponding even/odd pair generators.
    }
    \label{fig:tb_hamiltonian_dataflow}
\end{figure}

In addition to the full workflow, we implemented a stand-alone Hamiltonian-generation kernel to expose 
the peak throughput of the Hamiltonian evaluation logic independent of diagonalisation. 
Removing the diagonalisation stage, which occupies a substantial fraction of the available FPGA resources 
(see Sec.~\ref{subsec:hardware_resource}),
allows the Hamiltonian evaluation hardware to be replicated to increase throughput. In this configuration, 
pair generation and Hamiltonian-element evaluation are duplicated and operate in parallel on disjoint subsets 
of orbital pairs. The resulting element streams are merged and forwarded to the output interface. 
The corresponding dataflow graph is shown in Fig.~\ref{fig:tb_hamiltonian_dataflow}, 
illustrating the pipelined stages for pair generation and Hamiltonian evaluation.

Taken together, the design realises a fully streaming, on-device electronic-structure 
workflow in which successive orbital pairs and molecular geometries are processed without 
host intervention. This architectural structure underpins the performance characteristics 
examined in the following sections.

\subsection{Execution-Time Scaling of FPGA EHT and DFTB0 workflows}
\label{subsec:fpga_runtime_scaling}

\begin{figure}[h]
    \centering
    \includegraphics[scale=1]{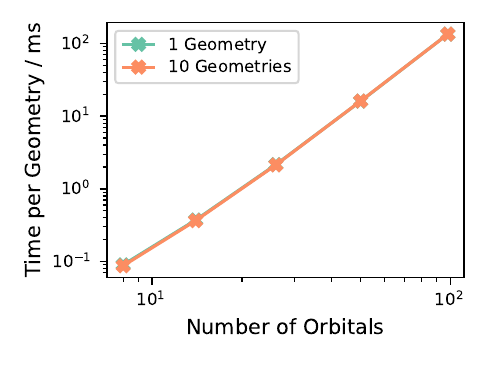}
    \caption{
        Execution time per processed geometry for the EHT module on the FPGA as a
        function of the number of atomic orbitals.
    }
    \label{fig:eht_benchmark}
\end{figure}

\begin{figure}[h]
    \centering
    \includegraphics[scale=1]{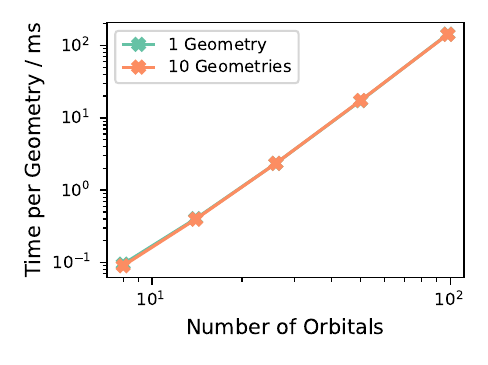}
    \caption{
        Execution time per processed geometry for the DFTB0 module on the FPGA as a
        function of the number of atomic orbitals.
    }
    \label{fig:dftb0_benchmark}
\end{figure}

The execution times per processed geometry for the FPGA implementations of the
EHT and DFTB0 workflows are shown in 
Fig.~\ref{fig:eht_benchmark} and Fig.~\ref{fig:dftb0_benchmark}, respectively,
as a function of the number of atomic orbitals.
The benchmarks are performed on a series of linear alkanes from methane (\hc{}{4}) up to
\textit{n}-hexadecane (\hc{16}{34}). The generation of the molecular test set and the FPGA
benchmarking protocol are described in Sec.~\ref{subsec:benchmarking_protocol}.
FPGA resource utilisation for the EHT and DFTB0 designs across these test molecules is
summarised in Sec.~\ref{subsec:hardware_resource}.
All reported execution times are averaged over 50 randomly generated geometries. 
For a fixed geometry, the FPGA runtime is fully deterministic and repeatable; 
the observed spread across geometries arises solely from geometry-dependent variations 
in the number of Jacobi sweeps required to reach the convergence criterion.

Over the considered system-size range, the execution time scales close to
$N_\mathrm{orb}^3$, indicating that the overall runtime is dominated by the
cyclic Jacobi eigensolver used in the hardware design.

The per-geometry execution times obtained from the single-geometry and
ten-geometry configurations are nearly identical.
This reflects the fact that the diagonalisation stage effectively serialises
the workflow and limits the degree of overlap between successive geometries.
As a result, batching multiple geometries does not significantly reduce the
average execution time per geometry when the full workflow is executed.

In our implementation, the Hamiltonian-construction stage contributes less 
to the total runtime for DFTB0 than for EHT. The overall DFTB0 workflow is 
nevertheless slightly slower because the resulting matrices typically 
require more Jacobi sweeps for convergence for the same set of molecular systems.

\begin{figure}[h]
    \centering
    \includegraphics[scale=1]{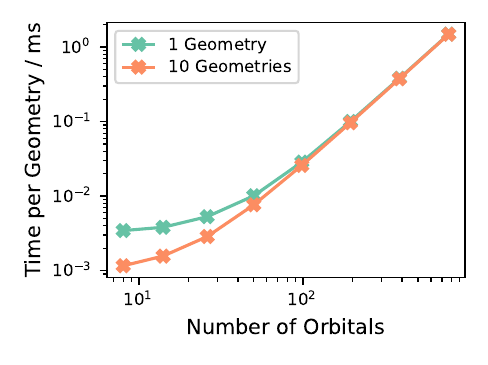}
    \caption{
        Execution time per processed geometry for the stand-alone 
        DFTB0 Hamiltonian generation module on the FPGA as a
        function of the number of atomic orbitals.
    }
    \label{fig:dftb0_hamiltonian_benchmark}
\end{figure}

The execution times per processed geometry for the DFTB0 Hamiltonian-generation kernel
are shown in Fig.~\ref{fig:dftb0_hamiltonian_benchmark}.
The benchmarks use the same series of linear alkanes as in the full-workflow tests,
extended up to \textit{n}-octacosahectane (\hc{128}{258}).
The generation of the molecular test set and the FPGA benchmarking protocol are described 
in Sec.~\ref{subsec:benchmarking_protocol}.
FPGA resource utilisation is summarised in Sec.~\ref{subsec:hardware_resource}.
The reported values correspond to the time required to generate the
upper-triangular part of the Hamiltonian for a single molecular geometry
and are deterministic for a given system size.
The runtime exhibits an approximately
quadratic scaling with the number of atomic orbitals, consistent with the
pairwise structure of the underlying algorithm.
For small systems, pipeline start-up latency constitutes a noticeable fraction
of the total runtime.
As the system size increases, this overhead becomes negligible and the steady-
state throughput of the pipelined design dominates.
In the ten-geometry configuration, partial overlap between coordinate loading
and Hamiltonian evaluation further reduces the average execution time per
geometry, although the benefit diminishes for larger systems where data loading
scales linearly with system size.

Overall, these results show that, for the full FPGA EHT and DFTB0 workflows, 
the execution time per processed geometry scales cubically across all the 
considered molecular systems and is largely insensitive to batching in the one-geometry 
\emph{vs.} ten-geometry configurations, consistent with the cyclic Jacobi diagonalisation 
dominating end-to-end runtime and effectively serialising the streaming task graph. 
For the stand-alone DFTB0 Hamiltonian-generation kernel, the runtime scales approximately 
quadratically as expected, with only minor small-system start-up overhead 
that can be partially mitigated through geometry batching.
% Consequently, the observed scaling directly reflects the dominant computational
% kernels, with cubic behaviour for the full workflows governed by the cyclic
% Jacobi diagonalisation and quadratic behaviour for the stand-alone Hamiltonian
% generation.

\subsection{Comparison With C++ Reference Implementation}
\label{subsec:fpga_cpu_comparison}

Tables~\ref{tab:eht_runtime_comparison}, \ref{tab:dftb0_runtime_comparison}, and
\ref{tab:dftb0_hamiltonian_runtime_comparison} summarise the execution-time statistics 
for the FPGA and C++ implementations of the EHT and DFTB0 workflows, 
as well as for the stand-alone DFTB0 Hamiltonian-generation kernel.
For each workflow, the tables report the mean execution time per processed
geometry together with the relative run-to-run variability quantified as the
interquartile-range-to-mean ratio (IQR/mean).
Results are shown for both single-geometry and ten-geometry configurations.
For the C++ reference implementation, timing values were obtained from the same
50 geometries used for the FPGA measurements, with 100 repeated runs per geometry 
(see Sec.~\ref{subsec:benchmarking_protocol}).
The reported statistics are computed over all 5000 individual timings and therefore 
reflect both geometry-dependent variation and run-to-run fluctuations. 
For the FPGA, run-to-run variability is absent for fixed geometries, and the reported 
spread arises solely from geometry-dependent effects
(see Sec.~\ref{subsec:fpga_runtime_scaling}).

The stand-alone DFTB0 Hamiltonian-generation kernel shows that the FPGA design 
can provide a clear performance benefit for workloads dominated by regular, 
streaming-friendly arithmetic. 
Here, the implementation can fully exploit a deeply pipelined datapath and 
replication of compute units, so that many pairwise contributions are evaluated concurrently 
while new data are accepted every cycle. This design sustains a high, 
geometry-independent throughput.
As shown in Table~\ref{tab:dftb0_hamiltonian_runtime_comparison}, the FPGA kernel 
becomes faster than the C++ reference for moderate system sizes and achieves a speed-up 
exceeding fourfold for the larger molecules considered. 
In addition, the FPGA execution time is deterministic, with IQR/mean = 0.0\,\% across all 
reported cases in Table~\ref{tab:dftb0_hamiltonian_runtime_comparison}.
This advantage is consistent with the underlying algorithmic structure of Hamiltonian construction, 
which can be mapped efficiently to parallel hardware pipelines and executed 
with minimal branching and synchronisation, allowing the design to translate available 
on-chip parallelism directly into reduced wall-clock time.

Notice that for the full EHT and DFTB0 workflows, the FPGA implementation exhibits longer 
execution times than the C++ reference across the considered system sizes. 
This behaviour is primarily associated with the diagonalisation stage, which dominates the 
overall runtime in the FPGA design. The hardware implementation employs a cyclic Jacobi eigensolver, 
which offers regular control flow but requires more floating-point operations than 
the QR- or divide-and-conquer-based solvers used in established CPU libraries.
Nevertheless, the FPGA workflows retain the deterministic execution behaviour 
established in Sec.~\ref{subsec:fpga_runtime_scaling}.

Overall, these results highlight a clear distinction between algorithmic
components that are currently limited by eigenvalue solvers and those that are
well suited to streaming, data-parallel hardware execution.
While the full EHT and DFTB0 workflows are dominated by diagonalisation and
therefore favour highly optimised CPU solvers, the Hamiltonian-generation stage
already demonstrates a substantial performance advantage on the FPGA.
This suggests that advances in hardware-friendly diagonalisation schemes or
hybrid execution strategies could narrow or even eliminate the performance gap
for complete workflows, while retaining, and potentially extending, the strong
throughput advantages observed for individual kernels.

\subsection{Energy and power characteristics}
\label{subsec:energy_and_power_characteristics}
One of the biggest advantages of FPGA computing is its energy efficiency.
To quantify this aspect for the considered workloads,
power consumption was measured using the largest system sizes considered 
for each kernel: \hc{16}{34} for the full EHT and DFTB0 workflows, 
and \hc{128}{258} for the stand-alone DFTB0 Hamiltonian-generation kernel.
Table~\ref{tab:power_and_energy} reports the averaged power draw and resulting energy 
per processed geometry.
The power and energy measurement protocols are described in Sec.~\ref{subsec:energy_measurement}.
The associated measurement datasets are provided in the Supplementary Information.

For the full EHT and DFTB0 workflows, the FPGA exhibits a substantially lower
instantaneous power draw than the CPU, remaining below 0.4\,W for both methods.
However, the longer execution times of the complete workflows on the FPGA
(\emph{ca.}~140\,ms per geometry) lead to a higher energy per geometry than the
CPU when system-level baseline power is excluded.
This behaviour is consistent with the performance results discussed in
Section~\ref{subsec:fpga_cpu_comparison}.
When system-level baseline power consumption is included, the energy per
geometry of the FPGA workflows becomes comparable to, or lower than, that of the
CPU, owing to the large disparity in instantaneous power draw between the two
platforms.

A qualitatively different behaviour is observed for the stand-alone DFTB0
Hamiltonian-generation kernel.
Here, the FPGA combines a low power draw with a shorter execution time than the
CPU, resulting in a substantially lower energy per processed geometry.
As shown in Table~\ref{tab:power_and_energy}, the FPGA requires less than
1\,mJ per geometry, whereas the CPU requires several hundred millijoules,
depending on the system-level baseline power consumption.
This result illustrates that for computational kernels with regular control
flow and high data parallelism, FPGA implementations can offer clear advantages
in both throughput and energy efficiency.

\section{Discussion}\label{sec:discussion}

Targeting high-throughput, multi-geometry workloads, this study shows that semi-empirical 
electronic-structure workflows can be organised as a hardware-native
streaming pipeline, where Hamiltonian construction and subsequent linear-algebra processing 
form a deterministic dataflow. 
The results demonstrate that pairwise Hamiltonian construction is naturally 
streaming-friendly and maps efficiently to deep pipelines, whereas end-to-end 
throughput and energy are currently governed by dense diagonalisation. 
We view this study as an architectural proof-of-concept rather than a final 
performance optimum, and the main takeaway is that adopting a more hardware-efficient eigensolver and/or increasing solver-level parallelism is the most direct route to improving both throughput and energy efficiency for the full workflow.

Several development paths could alleviate the current bottlenecks. 
Specialised FPGA eigensolver designs can better exploit available parallelism than 
generic IP cores or library implementations, and prior work has reported substantial
speedups for Jacobi- and Hestenes--Jacobi-type approaches.~\cite{bravo_high_2015,shi_accelerating_2020,zhou_field_2024}
In addition, throughput can be increased by instantiating multiple independent eigensolvers 
to process different geometries concurrently, even if each solver remains largely sequential. 
A heterogeneous workflow is also attractive, in which the FPGA constructs the Hamiltonian 
and diagonalisation is executed on integrated or attached CPU cores using optimised BLAS/LAPACK routines.
Such partitioning is naturally compatible with modern FPGA platforms that integrate programmable
logic with application-class CPUs.
Finally, porting the design to newer FPGA generations with higher clock
frequencies, greater DSP and on-chip memory resources (including UltraRAM)
would support deeper parallelism and larger systems. Together, these options
outline a clear route to improved throughput and scalability for FPGA-based
electronic-structure calculations.

Looking forward, several extensions could expand the capabilities of the FPGA-accelerated 
workflow. Implementing analytic nuclear gradients would enable geometry optimisation 
and molecular dynamics, while incorporating self-consistent charge (SCC) DFTB would 
improve accuracy at the expense of additional iterations, broadening the method's applicability 
for large screening campaigns. 
Excited-state calculations, for instance \emph{via} time-dependent DFTB, also appear
feasible on FPGA platforms using iterative solvers such as Lanczos.~\cite{choy_high_2012,sgherzi_solving_2021}
This would, however, require additional response-specific operators beyond the
ground-state Hamiltonian and therefore modifications to the current pipeline.
Together, these developments would move towards complete FPGA-accelerated semi-empirical
toolchains for screening, optimisation, dynamics, and spectroscopic observables.

Beyond methodological extensions, FPGAs offer an inherently energy-efficient
execution model for streaming workloads. 
This aligns with the broader observation that the fine-grained parallelism 
and low control overhead can yield substantially reduced energy consumption 
per operation on FPGAs.~\cite{qasaimeh_comparing_2019,nguyen_fpgabased_2022,choppali_sudarshan_greenfpga_2024}
In high-throughput settings, where many small or moderately sized matrix problems must be solved 
repeatedly, such characteristics have the potential not only to accelerate computation 
but also to lessen the overall energy footprint of electronic-structure calculations. 
Together with the architectural developments outlined above, these characteristics point towards 
a route for sustainable, high-throughput electronic-structure simulations in which performance 
and energy efficiency go hand in hand.

\section{Methods}\label{sec:methods}

\subsection{FPGA Workflow} \label{subsec:fpga_workflow}

\subsubsection{Devices, tools and clocking} \label{subsubsec:devices_tools_clocking}

Our design was realised through high-level synthesis on a Digilent Arty A7-100T 
development board equipped with a Xilinx Artix-7 FPGA (XC7A100TCSG324-1), 
a mid-range, low-power FPGA optimised for embedded applications. 
High-level synthesis was performed using Vitis HLS v2024.2, and hardware integration 
and bit-stream generation were carried out in Vivado v2024.2. The design was synthesised 
for a target clock frequency of 100\,MHz. All kernels were generated through
out-of-context synthesis for fixed molecular sizes, so that loop bounds and memory
dimensions were known at synthesis time. 
Accordingly, we generated a separate FPGA bitstream for each molecular size 
and for each streaming configuration used in the benchmarks (single-geometry and ten-geometry workflows).
A MicroBlaze soft-core processor was used
to control the HLS kernel, handle data exchange with the host computer \emph{via} an 
AXI UART Lite intellectual property (IP) core (v2.0) for debugging, and perform
timing measurements through an AXI Timer IP (v2.0). On-board supply voltages and currents 
were monitored using an XADC Wizard IP (v3.3), which was configured for power estimation 
during FPGA runs. The reference data for testbenches are generated using
PySCF~\cite{sun_recent_2020} and DFTB+~\cite{hourahine_dftb_2020}.

The MicroBlaze firmware was developed in C using Vitis v2024.2. Diagonalisation was 
implemented with the Jacobi eigenvalue solver from the AMD Vitis Solver Library v2025.1, 
adapted for the Artix-7 architecture. Due to some issues in the library, the Jacobi rotations 
cannot be parallelised across multiple compute units on our Artix-7, which limits the 
throughput of the diagonalisation stage. The complete project files and source code 
used in this work are provided in the Supplementary Information.

\subsubsection{General Design Principles}
\label{subsubsec:general_design_principles}

\paragraph{Loop Structuring.}
Loops dominate the computational kernels of tight-binding methods and strongly influence 
the achievable level of pipelining in HLS-based FPGA designs.
For efficient synthesis, loops should be flattened
whenever possible, avoiding deep nesting, and should have statically bounded
trip counts. In our workflow, these bounds are made static by making the designs
molecule specific. This allows the HLS tool to optimise loop scheduling and 
control flow without introducing runtime logic. This approach can be generalised 
to designs that accommodate molecules up to prescribed maximum sizes. Furthermore, 
we compute pairs of matrix indices $(i, j)$ in a dedicated pair-generation stage 
at the beginning of the pipeline. By doing so, the nested double loop over orbital
indices is transformed into a flat stream of index pairs that is passed to all
downstream stages, thereby avoiding nested loops.

\paragraph{Pipelining at All Levels.}
Pipelining is ubiquitous; CPUs and GPUs employ it internally, but on FPGAs it
is a first-class, designer-controlled mechanism. Suppose a task consists of $N$
stages, each taking $l$ cycles to complete. If the stages run sequentially, the
per-item service time is $N\, l$. In a pipeline, each stage can immediately
accept the next input once it completes its work. Thus, a new iteration begins 
every $l$ cycles, which is called the initiation interval ($\mathrm{II}$). 
The pipeline takes $L=N\, l$ cycles to fill and emit the first result, but 
produces subsequent results at a rate of one every $l$ cycles.

This concept is applied at multiple levels in our design.
At the operator level, arithmetic units that require several cycles to complete,
\emph{e.g.} floating-point addition and multiplication, are inferred as pipelined
units by the HLS tool. At the loop level, all stages that process pair data are 
scheduled to achieve $\mathrm{II} = 1$, enabling the production of one
Hamiltonian element per cycle once the pipeline is full. As the design accepts 
a sequence of geometries, each stage can begin processing geometry $t+1$ 
immediately after finishing geometry $t$. If stage $k$ requires $C_k$ cycles 
per geometry, then once the pipeline stabilises, the inter-geometry $\mathrm{II}$ 
becomes $\max_k C_k$, rather than $\sum_k C_k$ for sequential execution.

\paragraph{Arbitrary-Sized Data Types.}
Unlike CPUs/GPUs, which expose a small, predefined set of widths, FPGAs allow
bit-accurate widths for both data and addresses. This flexibility means on-chip 
storage and interconnect scale with the actual information content, and 
memory or stream widths can be matched to physical resources instead of wasting lanes. 
In our current design, we exploit this mainly for indices and addresses: 
orbital and atom indices, pair counters, and array addresses use minimal‐width 
arbitrary-precision unsigned-integer (\texttt{ap\_uint}) types, which minimise 
silicon area, reduce energy consumption, and can lower computation latency.

\subsection{Method-Specific Kernels}
\label{subsec:method_specific_kernels}

\subsubsection{Extended Hückel Theory}
\label{subsubsec:method_specific_eht}

Because the EHT Hamiltonian is constructed directly from the overlap matrix,
the only molecular integrals that need to be evaluated are the overlap
integrals. For Gaussian-type orbitals with low angular momentum (restricted
here to $l \leq 1$), it is feasible to implement the overlap integrals using
closed-form expressions, avoiding the need for recursive schemes. 
In our implementation, each orbital $\mu$ is characterised by an exponent $\alpha_{\mu}$,
a prefactor $d_{\mu}$, its centre $\vec{R}_{\mu}$, and its angular momentum $l \in \{s,p_x,p_y,p_z\}$.
For a pair $(\mu, \nu)$, the overlap integral between two s-type orbitals is given by
\begin{equation}
    S_{\mu \nu}^{s,s} = d_{\mu} d_{\nu} \left(\frac{\pi}{p_{\mu \nu}}\right)^{3/2} \exp\left(-\frac{\alpha_\mu \alpha_\nu}{p_{\mu \nu}} \|\vec{R}_{\mu \nu}\|^2\right)\,,
\end{equation}
where $p_{\mu \nu} = \alpha_\mu + \alpha_\nu$ and $\vec{R}_{\mu \nu} = \vec{R}_{\nu} - \vec{R}_{\mu}$ 
with the components $\left( R_{\mu \nu}^x, R_{\mu \nu}^y, R_{\mu \nu}^z \right)$.
The overlap integral between an s-type and a p-type orbital is given by
\begin{align}
    S_{\mu \nu}^{s,p_a} &= -\frac{\alpha_\mu}{p_{\mu \nu}} R_{\mu \nu}^a  S_{\mu \nu}^{s,s}\,, \\
    S_{\mu \nu}^{p_a,s} &=  \frac{\alpha_\nu}{p_{\mu \nu}} R_{\mu \nu}^a  S_{\mu \nu}^{s,s}\,,
\end{align}
for $a \in \{x, y, z\}$.
The overlap integral between two p-type orbitals is given by
\begin{equation}
    S_{\mu \nu}^{p_a,p_b} =
    \begin{cases}
        \left(\frac{1}{2p_{\mu \nu}} - \frac{\alpha_\mu \alpha_\nu}{p_{\mu \nu}^2}(R_{\mu \nu}^a)^2\right) S_{\mu \nu}^{s,s} & a = b \,, \\
        \left(-\frac{\alpha_\mu \alpha_\nu}{p_{\mu \nu}^2} R_{\mu \nu}^a R_{\mu \nu}^b\right) S_{\mu \nu}^{s,s} & a \neq b \,, \\
    \end{cases}
\end{equation}
for $a, b \in \{x, y, z\}$.
The required scaling factors and valence orbital energies are supplied as 
constants at synthesis time, allowing each Hamiltonian element to be computed 
immediately once its corresponding overlap integral is available.

\subsubsection{Non-Self-Consistent DFTB}
\label{subsubsec:method_specific_dftb0}

In DFTB0, Hamiltonian elements are obtained from pre-tabulated two-centre
integrals through interpolation followed by Slater–Koster transformations.  
The tabulated functions depend only on the interatomic distance and are stored
in on-chip block RAM, which provides deterministic, low-latency random access
well suited for lookup-based evaluation.

Each Slater--Koster table is defined on a regular radial grid. To evaluate the
integral at a distance $r$, a linear interpolation is performed. In its
standard form, the interpolated value is
\begin{equation*}
    y(r) = y(r_i) + \frac{r - r_i}{r_{i+1} - r_i} \left[ y(r_{i+1}) - y(r_i) \right]\,,
\end{equation*}
which requires reading two adjacent grid points. To avoid this on hardware, the
grid values are stored as tuples $(y_i, \Delta y_i)$ with
$\Delta y_i = y_{i+1} - y_i$. After computing the fractional position
$t = (r - r_i)/(r_{i+1} - r_i)$, the interpolation reduces to a single fused
multiply–add (FMA):
\begin{equation*}
    y(r) = y(r_i) + t \cdot \Delta y\,.
\end{equation*}
This representation requires only one memory read and one floating-point FMA
per lookup, enabling a fully pipelined evaluation stage that produces one
Hamiltonian element per cycle once the pipeline is filled.

Beyond the Hamiltonian elements, DFTB0 requires evaluating the repulsive pair
potential for the distinct atom pairs. The traditional DFTB parametrisation
defines this potential piecewise using different analytic forms in separate
distance regions, which leads to irregular control flow when implemented on an
FPGA. To maintain a uniform access pattern, we approximate the entire
repulsive potential by a single cubic spline on an equidistant grid. Each spline
segment is described by four coefficients $(c_0, c_1, c_2, c_3)$, stored in
on-chip memory. The value for a distance $r$ is evaluated using Horner’s rule,
\begin{equation*}
    V_{\mathrm{rep}}(r) = c_0 + r \left( c_1 + r \left( c_2 + r c_3 \right) \right)\,,
\end{equation*}
which maps to a short arithmetic pipeline composed of chained FMAs. 

Because the coefficients are read sequentially in our design, the spline evaluator 
produces one value every four cycles. 
Its inputs are generated by the pair-generation stage, which emits two streams:
one containing ordered orbital pairs for Hamiltonian evaluation, and one containing
the distinct atom pairs required for the repulsive potential. 

\subsection{Hardware resource characteristics}
\label{subsec:hardware_resource}

The designs were synthesised for the Artix-7 FPGA platform described in
Sec.~\ref{subsubsec:devices_tools_clocking}.
The resource utilisation of the EHT and DFTB0 computational modules is summarised in 
Tables~\ref{tab:eht_resources} and \ref{tab:dftb0_resources}, respectively. 
Each table reports the number of look-up tables (LUT), flip-flops (FF), block RAMs (BRAM), 
and digital signal processing slices (DSP) as absolute counts and percentages 
of available resources. 
These values also include the small amount of logic required for debug 
intellectual property (IP) cores.

Across the considered molecular sizes, LUT and FF counts change only slightly, whereas
BRAM usage increases more strongly with system size. This trend reflects the growing
on-chip storage requirements for the Hamiltonian and auxiliary matrices used during
diagonalisation.
Comparing the two methods, the DFTB0 module uses more LUT, FF, and BRAM than the
EHT module, primarily due to control logic and on-chip storage for Slater--Koster parameter
access and interpolation. Conversely, DSP usage is lower, as no explicit integral evaluation
is required. The single-geometry and ten-geometry configurations differ only marginally
in logic utilisation.

The hardware resources required by the stand-alone DFTB0 Hamiltonian-generation kernel
are summarised in Table~\ref{tab:dftb0_hamiltonian_resources}. 
As before, these values include the small amount of logic associated with debug IP cores. 
Across the tested molecular sizes, LUT and FF usage varies only slightly, 
while BRAM usage shows step-like changes due to automatic memory mapping 
(distributed LUT RAM \emph{versus} inferred BRAM) as array sizes cross implementation
thresholds. Overall BRAM growth remains moderate because the Slater--Koster parameter
tables dominate the memory footprint and are independent of system size. DSP utilisation
is essentially constant across all designs. The single-geometry and ten-geometry
configurations differ only marginally in resource utilisation.

\subsection{Benchmarking Protocol}
\label{subsec:benchmarking_protocol}

For performance evaluation, the FPGA implementation on the Artix-7 FPGA was compared 
against a molecule-specific C++ implementation executed on a single core of an Intel\textsuperscript{\textregistered} Xeon\textsuperscript{\textregistered}
E5-2660~v3 CPU running at 2.60\,GHz.
This processor is representative of server-class CPUs used in high-performance 
computing, and both devices originate from a comparable technology era (2012–2014), 
allowing a balanced assessment of per-device efficiency. 
As test systems, a series of linear alkanes from methane (\hc{}{4}) up to 
\textit{n}-hexadecane (\hc{16}{34}) was considered for the full EHT and DFTB0 workflows, 
and extended the stand-alone DFTB0 Hamiltonian-generation kernel benchmarks up to 
\textit{n}-octacosahectane (\hc{128}{258}). 

For each molecule, a single reference geometry was used as the starting point. 
From this reference structure, 50 geometry sets were generated by applying 
random perturbations to atomic positions together with a random rigid rotation 
of the entire molecule. Each geometry set contained either one perturbed structure 
(for the single-geometry workflow) or ten independently perturbed structures 
(for the ten-geometry workflow). The same geometry sets were supplied to both 
FPGA and CPU implementations to ensure a fair comparison between platforms.

Timing on the FPGA was measured using the AXI Timer IP controlled by the 
MicroBlaze processor. The timer was started immediately before launching the HLS kernel 
and stopped once all streamed geometries had been processed. UART communication and 
any debug output were kept outside the timed region. On the CPU, timings were collected 
using \texttt{std::chrono} around the core Hamiltonian-construction and diagonalisation routines.
Dense diagonalisation was performed using Eigen's \texttt{Eigen::SelfAdjointEigenSolver}, 
which reduces the matrix to tridiagonal form and applies an implicit symmetric QR algorithm
with Wilkinson shift.
The C++ codes were compiled using CMake in \texttt{Release} mode with \texttt{gcc/g++} 
and full optimisation (\texttt{-O3}).

For the complete EHT and DFTB0 workflows, each of the 50 geometry sets was evaluated 
once on the FPGA, yielding one deterministic timing per set apart from the data-dependent 
number of Jacobi sweeps in the diagonalisation stage. The corresponding CPU implementations 
were executed 100 times per geometry set, providing 5000 timings per molecule for 
statistical analysis. For the stand-alone DFTB0 Hamiltonian-generation kernel, the FPGA design
produced a single deterministic timing for each molecular size, whereas the CPU version was
evaluated for 50 geometry sets with 100 repetitions each (5000 timings). The reported timings 
correspond to averages over all measured runs for each molecule and workflow.

\subsection{Energy Measurement}
\label{subsec:energy_measurement}

Energy measurements were performed both on the FPGA and on the CPU over
extended time windows to obtain stable long-term averages. On the FPGA, power
was measured using the XADC Wizard IP configured in continuous-conversion mode
with 256-sample hardware averaging. The VCCINT rail was monitored \emph{via} the
built-in voltage channel, and the core supply current was obtained from the
VAUXP10/VAUXN10 differential input, which on the Arty A7 board is wired across
the shunt resistor of the core supply. XADC samples were read periodically
through the JTAG (Joint Test Action Group) interface at approximately 5\,ms intervals
during repeated execution of the workflow over roughly 300\,s. Instantaneous
power was obtained from the measured voltage and current values, and the mean
over all samples was reported.

CPU energy was measured using Intel's RAPL (Running Average Power Limit) interface, 
which exposes model-based package energy counters through the Linux powercap subsystem. 
The counters were sampled once per second over a period of approximately 300\,s 
during repeated CPU execution of the workflow to avoid counter wrap-around. 
Total package energy was obtained from the difference between the initial and final 
counter values, and average CPU power was computed by dividing this energy by the 
elapsed wall time. To distinguish load-dependent energy from background consumption, 
an additional idle RAPL measurement was performed, and both raw and baseline-subtracted 
average powers are reported.

\backmatter

\bmhead{Supplementary information}

Supplementary Software~1 contains a complete archive of the project files and
source code used in this work, including all HLS kernels, FPGA designs, 
MicroBlaze firmware, and benchmarking scripts. A static snapshot of the repository 
is provided as a \texttt{.zip} file accompanying the submission; 
the development version is available at
\url{https://github.com/mitric-lab/fpga_tight_binding}.

Supplementary Data~1 provides the timing and power measurement datasets used to
produce the results in the main text, supplied in \texttt{.csv} format.

\bmhead{Acknowledgements}
X. M. thanks Studienstiftung des deutschen Volkes for a fellowship.

\section*{Declarations}

\begin{itemize}
\item Funding: X. M. is funded by Studienstiftung des deutschen Volkes.
\item Competing interests: The authors declare no competing interests.
\item Ethics approval and consent to participate: Not applicable.
\item Consent for publication: Not applicable.
\item Data availability: All the datasets used in this work are available in the supplementary information.
\item Materials availability: Not applicable.
\item Code availability: All the source codes used in this work are available in the supplementary information.
\item Author contribution: X. M. contributed to software, formal analysis, investigation, data curation, writing -- original draft, and visualisation. R. M. contributed to conceptualisation, resources, supervision, project administration, and funding acquisition. Both authors contributed to methodology, validation, and writing -- review \& editing.
\end{itemize}

\noindent
If any of the sections are not relevant to your manuscript, please include the heading and write `Not applicable' for that section.

\bibliography{references}% common bib file

\clearpage

\section*{Tables}

\begin{table}[h]
    \centering
    \caption{
        Execution-time statistics for the FPGA and C++ implementations of the 
        EHT workflow.
        Reported values correspond to the mean execution time per processed geometry
        and the relative variability expressed as IQR/mean (in percent).
        Results are shown for single-geometry and ten-geometry configurations.
    }
    \label{tab:eht_runtime_comparison}
    \begin{tabular}{lrrrr}
        \toprule
                    & \multicolumn{2}{c}{C++} & \multicolumn{2}{c}{FPGA} \\
                    & mean / ms & IQR/mean    & mean / ms & IQR/mean     \\ \midrule
        \multicolumn{5}{l}{\textbf{Single-geometry design}}              \\
        \hc{}{4}    & 0.039 & 17.4\,\% &   0.090 &  9.2\,\% \\
        \hc{2}{6}   & 0.049 & 14.5\,\% &   0.373 &  7.0\,\% \\
        \hc{4}{10}  & 0.091 & 10.3\,\% &   2.159 &  6.6\,\% \\
        \hc{8}{18}  & 0.243 &  6.0\,\% &  16.086 &  3.6\,\% \\
        \hc{16}{34} & 1.084 &  4.1\,\% & 134.776 &  3.9\,\% \\
        \midrule
        \multicolumn{5}{l}{\textbf{Ten-geometry design}}                 \\
        \hc{}{4}    & 0.008 &  9.5\,\% &   0.087 &  3.2\,\% \\
        \hc{2}{6}   & 0.017 &  5.0\,\% &   0.366 &  2.8\,\% \\
        \hc{4}{10}  & 0.052 &  4.0\,\% &   2.134 &  1.8\,\% \\
        \hc{8}{18}  & 0.192 &  4.0\,\% &  16.165 &  1.6\,\% \\
        \hc{16}{34} & 0.953 &  1.3\,\% & 134.510 &  1.5\,\% \\
        \bottomrule
    \end{tabular}
\end{table}

\begin{table}[h]
    \centering
    \caption{
        Execution-time statistics for the FPGA and C++ implementations of the
        DFTB0 workflow.
        Reported values correspond to the mean execution time per processed geometry
        and the relative variability expressed as IQR/mean (in percent).
        Results are shown for single-geometry and ten-geometry configurations.
    }
    \label{tab:dftb0_runtime_comparison}
    \begin{tabular}{lrrrr}
        \toprule
                    & \multicolumn{2}{c}{C++} & \multicolumn{2}{c}{FPGA} \\
                    & mean / ms & IQR/mean    & mean / ms & IQR/mean     \\ \midrule
        \multicolumn{5}{l}{\textbf{Single-geometry design}}              \\
        \hc{}{4}    & 0.019 & 21.6\,\% &   0.095 &  7.0\,\% \\
        \hc{2}{6}   & 0.033 & 13.4\,\% &   0.404 &  4.8\,\% \\
        \hc{4}{10}  & 0.075 &  6.7\,\% &   2.354 &  4.5\,\% \\
        \hc{8}{18}  & 0.245 &  3.8\,\% &  17.257 &  3.9\,\% \\
        \hc{16}{34} & 1.037 &  3.8\,\% & 144.072 &  3.6\,\% \\
        \midrule
        \multicolumn{5}{l}{\textbf{Ten-geometry design}}                 \\
        \hc{}{4}    & 0.007 &  7.3\,\% &   0.090 &  2.3\,\% \\
        \hc{2}{6}   & 0.017 &  3.4\,\% &   0.397 &  1.9\,\% \\
        \hc{4}{10}  & 0.052 &  2.0\,\% &   2.353 &  1.6\,\% \\
        \hc{8}{18}  & 0.207 &  3.8\,\% &  17.401 &  1.6\,\% \\
        \hc{16}{34} & 0.946 &  3.1\,\% & 142.787 &  1.2\,\% \\
        \bottomrule
    \end{tabular}
\end{table}

\begin{table}[h]
    \centering
    \caption{
        Execution-time statistics for the FPGA and C++ implementations of the
        stand-alone DFTB0 Hamiltonian-generation kernel.
        Reported values correspond to the mean execution time per processed geometry
        and the relative variability expressed as IQR/mean (in percent).
        Results are shown for single-geometry and ten-geometry configurations.
    }
    \label{tab:dftb0_hamiltonian_runtime_comparison}
    \begin{tabular}{lrrrr}
        \toprule
                      & \multicolumn{2}{c}{C++} & \multicolumn{2}{c}{FPGA} \\
                      & mean / ms & IQR/mean    & mean / ms & IQR/mean     \\ \midrule
        \multicolumn{5}{l}{\textbf{Single-geometry design}}              \\
        \hc{}{4}      & 0.0012 &  6.8\,\% & 0.0034 & 0.0\,\% \\
        \hc{2}{6}     & 0.0030 &  3.7\,\% & 0.0038 & 0.0\,\% \\
        \hc{4}{10}    & 0.0087 &  2.0\,\% & 0.0053 & 0.0\,\% \\
        \hc{8}{18}    & 0.0324 &  1.4\,\% & 0.0101 & 0.0\,\% \\
        \hc{16}{34}   & 0.1214 &  1.2\,\% & 0.0286 & 0.0\,\% \\
        \hc{32}{66}   & 0.4628 &  1.7\,\% & 0.1004 & 0.0\,\% \\
        \hc{64}{130}  & 1.8067 &  3.8\,\% & 0.3823 & 0.0\,\% \\
        \hc{128}{258} & 6.8369 &  2.7\,\% & 1.4986 & 0.0\,\% \\
        \midrule
        \multicolumn{5}{l}{\textbf{Ten-geometry design}}                 \\
        \hc{}{4}      & 0.0006 &  2.8\,\% & 0.0012 & 0.0\,\% \\
        \hc{2}{6}     & 0.0020 &  1.1\,\% & 0.0016 & 0.0\,\% \\
        \hc{4}{10}    & 0.0068 &  0.6\,\% & 0.0029 & 0.0\,\% \\
        \hc{8}{18}    & 0.0279 &  0.7\,\% & 0.0076 & 0.0\,\% \\
        \hc{16}{34}   & 0.1157 &  3.4\,\% & 0.0258 & 0.0\,\% \\
        \hc{32}{66}   & 0.4483 &  2.1\,\% & 0.0967 & 0.0\,\% \\
        \hc{64}{130}  & 1.6572 &  2.4\,\% & 0.3767 & 0.0\,\% \\
        \hc{128}{258} & 6.4884 &  2.9\,\% & 1.4897 & 0.0\,\% \\
        \bottomrule
    \end{tabular}
\end{table}

\begin{table}[h]
    \centering
    \caption{
        Power draw and energy per processed geometry for the FPGA and C++
        implementations on a CPU.
        Values are reported for the ten-geometry configuration.
        CPU results are shown both with and without system-level baseline.
    }
    \label{tab:power_and_energy}
    
    \begin{tabular}{lrrr}
        \toprule
         & FPGA  & CPU (with baseline) & CPU (baseline excluded) \\ \midrule
        \multicolumn{4}{l}{\textbf{Average power / W}} \\
        EHT (\hc{16}{34})                 & 0.351 & 65.1 & 26.4 \\
        DFTB0 (\hc{16}{34})               & 0.379 & 65.7 & 27.0 \\
        DFTB0 Hamiltonian (\hc{128}{258}) & 0.469 & 64.1 & 25.3 \\
        \midrule
        \multicolumn{4}{l}{\textbf{Energy per geometry / mJ}} \\
        EHT (\hc{16}{34})                 & 47.2  & 62.1 & 25.2 \\
        DFTB0 (\hc{16}{34})               & 54.1  & 62.2 & 25.5 \\
        DFTB0 Hamiltonian (\hc{128}{258}) & 0.699 &  416 &  164 \\
        \bottomrule
    \end{tabular}
\end{table}

\begin{table}[h]
    \centering
    \caption{
        FPGA resource utilisation of the EHT hardware design for different
        test molecules. The upper and lower blocks correspond to the single-geometry
        and ten-geometry configurations, respectively.
        BRAM counts refer to 18\,kbit memory blocks.
    }
    \label{tab:eht_resources}
    \begin{tabular}{lrrrr}
        \toprule
                    & LUT            & FF             & BRAM         & DSP          \\ \midrule
        \multicolumn{5}{l}{\textbf{Single-geometry design}}                         \\
        \hc{}{4}    & 17618 (28\,\%) & 20222 (16\,\%) &  39 (14\,\%) & 103 (43\,\%) \\
        \hc{2}{6}   & 18282 (29\,\%) & 20613 (16\,\%) &  48 (18\,\%) & 103 (43\,\%) \\
        \hc{4}{10}  & 18792 (30\,\%) & 21346 (17\,\%) &  54 (20\,\%) & 108 (45\,\%) \\
        \hc{8}{18}  & 19889 (31\,\%) & 23044 (18\,\%) &  76 (28\,\%) & 108 (45\,\%) \\
        \hc{16}{34} & 20521 (32\,\%) & 26102 (21\,\%) & 158 (59\,\%) & 121 (50\,\%) \\
        \midrule
        \multicolumn{5}{l}{\textbf{Ten-geometry design}}                            \\
        \hc{}{4}    & 17637 (28\,\%) & 20220 (16\,\%) &  39 (14\,\%) & 103 (43\,\%) \\
        \hc{2}{6}   & 18312 (29\,\%) & 20616 (16\,\%) &  48 (18\,\%) & 103 (43\,\%) \\
        \hc{4}{10}  & 18844 (30\,\%) & 21389 (17\,\%) &  54 (20\,\%) & 108 (45\,\%) \\
        \hc{8}{18}  & 19949 (31\,\%) & 23049 (18\,\%) &  76 (28\,\%) & 108 (45\,\%) \\
        \hc{16}{34} & 20572 (32\,\%) & 26132 (21\,\%) & 158 (59\,\%) & 122 (51\,\%) \\
        \bottomrule
    \end{tabular}
\end{table}

\begin{table}[h]
    \centering
    \caption{
        FPGA resource utilisation of the DFTB0 hardware design for different
        test molecules. The upper and lower blocks correspond to the single-geometry
        and ten-geometry configurations, respectively.
        BRAM counts refer to 18\,kbit memory blocks.
    }
    \label{tab:dftb0_resources}
    \begin{tabular}{lrrrr}
        \toprule
                    & LUT            & FF             & BRAM         & DSP          \\ \midrule
        \multicolumn{5}{l}{\textbf{Single-geometry design}}                         \\
        \hc{}{4}    & 21380 (34\,\%) & 23801 (19\,\%) &  85 (31\,\%) &  90 (38\,\%) \\
        \hc{2}{6}   & 22051 (35\,\%) & 24193 (19\,\%) &  94 (35\,\%) &  90 (38\,\%) \\
        \hc{4}{10}  & 22340 (35\,\%) & 24503 (19\,\%) & 112 (41\,\%) &  95 (40\,\%) \\
        \hc{8}{18}  & 23541 (37\,\%) & 26203 (21\,\%) & 134 (50\,\%) &  95 (40\,\%) \\
        \hc{16}{34} & 24238 (38\,\%) & 29312 (23\,\%) & 216 (80\,\%) & 108 (45\,\%) \\
        \midrule
        \multicolumn{5}{l}{\textbf{Ten-geometry design}}                            \\
        \hc{}{4}    & 21535 (34\,\%) & 23863 (19\,\%) &  85 (31\,\%) &  90 (38\,\%) \\
        \hc{2}{6}   & 22195 (35\,\%) & 24252 (19\,\%) &  94 (35\,\%) &  90 (38\,\%) \\
        \hc{4}{10}  & 22484 (35\,\%) & 24565 (19\,\%) & 112 (41\,\%) &  95 (40\,\%) \\
        \hc{8}{18}  & 23685 (37\,\%) & 26268 (21\,\%) & 134 (50\,\%) &  95 (40\,\%) \\
        \hc{16}{34} & 24391 (38\,\%) & 29381 (23\,\%) & 216 (80\,\%) & 108 (45\,\%) \\
        \bottomrule
    \end{tabular}
\end{table}

\begin{table}[h]
    \centering
    \caption{
        FPGA resource utilisation of the DFTB0 Hamiltonian-generation hardware design
        for different test molecules. The upper and lower blocks correspond to the
        single-geometry and ten-geometry configurations, respectively.
        BRAM counts refer to 18\,kbit memory blocks.
    }
    \label{tab:dftb0_hamiltonian_resources}
    \begin{tabular}{lrrrr}
        \toprule
                      & LUT            & FF             & BRAM         & DSP          \\ \midrule
        \multicolumn{5}{l}{\textbf{Single-geometry design}}                           \\
        \hc{}{4}      & 16102 (25\,\%) & 16567 (13\,\%) & 102 (38\,\%) & 105 (44\,\%) \\
        \hc{2}{6}     & 16147 (25\,\%) & 16578 (13\,\%) & 102 (38\,\%) & 105 (44\,\%) \\
        \hc{4}{10}    & 15912 (25\,\%) & 15859 (13\,\%) & 120 (44\,\%) & 105 (44\,\%) \\
        \hc{8}{18}    & 15949 (25\,\%) & 15890 (13\,\%) & 120 (44\,\%) & 105 (44\,\%) \\
        \hc{16}{34}   & 16003 (25\,\%) & 15919 (13\,\%) & 120 (44\,\%) & 105 (44\,\%) \\
        \hc{32}{66}   & 16030 (25\,\%) & 15916 (13\,\%) & 122 (45\,\%) & 105 (44\,\%) \\
        \hc{64}{130}  & 16081 (25\,\%) & 15937 (13\,\%) & 124 (46\,\%) & 105 (44\,\%) \\
        \hc{128}{258} & 16131 (25\,\%) & 15987 (13\,\%) & 124 (46\,\%) & 105 (44\,\%) \\
        \midrule
        \multicolumn{5}{l}{\textbf{Ten-geometry design}}                              \\
        \hc{}{4}      & 16143 (25\,\%) & 16627 (13\,\%) & 102 (38\,\%) & 105 (44\,\%) \\
        \hc{2}{6}     & 16174 (26\,\%) & 16637 (13\,\%) & 102 (38\,\%) & 105 (44\,\%) \\
        \hc{4}{10}    & 15962 (25\,\%) & 15924 (13\,\%) & 120 (44\,\%) & 105 (44\,\%) \\
        \hc{8}{18}    & 15989 (25\,\%) & 15957 (13\,\%) & 120 (44\,\%) & 105 (44\,\%) \\
        \hc{16}{34}   & 16060 (25\,\%) & 15994 (13\,\%) & 120 (44\,\%) & 105 (44\,\%) \\
        \hc{32}{66}   & 16086 (25\,\%) & 15975 (13\,\%) & 122 (45\,\%) & 106 (44\,\%) \\
        \hc{64}{130}  & 16122 (25\,\%) & 15997 (13\,\%) & 124 (46\,\%) & 106 (44\,\%) \\
        \hc{128}{258} & 16171 (26\,\%) & 16035 (13\,\%) & 124 (46\,\%) & 106 (44\,\%) \\
        \bottomrule
    \end{tabular}
\end{table}

\clearpage

\section*{Figures}

\end{document}